\documentclass[aps,prl,10pt,twocolumn,floatfix,showpacs,superscriptaddress,longbibliography]{revtex4-1}

\usepackage{graphicx,amsmath,amsfonts,bm}
\usepackage[colorlinks=true, citecolor=blue, linkcolor=blue, urlcolor=blue]{hyperref}
\usepackage[all]{hypcap}
\usepackage{color}
\begin{document}
\title{Fractional charge oscillations in quantum spin Hall quantum dots}
\author{N. Traverso Ziani}
\email{niccolo.traverso@physik.uni-wuerzburg.de}
\affiliation{Institute of Theoretical Physics and Astrophysics, University of W\"urzburg, 97074 W\"urzburg, Germany}
\author{C. Fleckenstein}
\email{christoph.fleckenstein@physik.uni-wuerzburg.de}
\affiliation{Institute of Theoretical Physics and Astrophysics, University of W\"urzburg, 97074 W\"urzburg, Germany}
\author{G. Dolcetto}
\affiliation{Physics and Materials Science Research Unit, University of Luxembourg, L-1511 Luxembourg}
\author{B. Trauzettel}
\affiliation{Institute of Theoretical Physics and Astrophysics, University of W\"urzburg, 97074 W\"urzburg, Germany}
\begin{abstract}
We show that correlated two-particle backscattering can induce fractional charge oscillations in a quantum dot built at the edge of a two-dimensional topological insulator by means of magnetic barriers. The result nicely complements recent works where the fractional oscillations were obtained employing of semiclassical treatments. Moreover, since by rotating the magnetization of the barriers a fractional charge can be trapped in the dot via the Jackiw-Rebbi mechanism, the system we analyze offers the opportunity to study the interplay between this noninteracting charge fractionalization and the fractionalization due to two-particle backscattering. In this context, we demonstrate that the number of fractional oscillations of the charge density depends on the magnetization angle. Finally, we address the renormalization induced by two-particle backscattering on the spin density, which is characterized by a dominant oscillation, sensitive to the Jackiw-Rebbi charge, with a wavelength twice as large as the charge density oscillations.
\end{abstract}
\maketitle
\section{Introduction}
\noindent As nanotechnology and material science advance, bringing phenomena that are peculiar to high energy physics down to the energy scale of condensed matter physics becomes feasible. Apart from the celebrated Anderson-Higgs mechanism in superconductors\cite{anderson}, a paradigmatic example is graphene. Its low energy properties are well described by Dirac-like Hamiltonians\cite{graphene1}: Klein tunneling has been theoretically predicted\cite{klein1} and experimentally observed\cite{klein2} and Zitterbewegung is believed to matter for the motion of its electrons\cite{zitter}. More recently, it has been shown that the chiral anomaly\cite{Adler,fuji,chiral1} is crucial in the understanding of the electromagnetic response of Weyl semimetals\cite{Volovik2003,Wan2011,Xu2011,Burkov2011a,Burkov2011b,ChaoXing,add1,add2,Liu2014,Xu2015a,Xu2015b,JXiong,add3} and the behavior of two-dimensional topological insulators in the presence of magnetic barriers\cite{realspace}. Another connection between high energy and condensed matter physics is charge fractionalization due to the Jackiw-Rebbi mechanism\cite{Jack}, which has been shown to play a role in polyacetylene\cite{ssh}. Fractional charges with charge $e/2$ have also been recently proposed to appear in carbon nanotubes under the influence of non-uniform strain and magnetic fields\cite{berg}. More general fractional charges, corresponding to complex solitons,\cite{general} are hosted by magnetically defined quantum dots defined at the edges of two dimensional topological insulators\cite{realspace,qi,dolc1,dolc2}, even in the presence of weak interactions\cite{dolc1,dolc2}.\\
A different type of charge fractionalization is known to take place in strongly interacting condensed matter systems. Apart from the well established fractional quantum Hall effect in two spatial dimensions\cite{fqhe}, in one dimension, the interplay between strong spin-orbit coupling and electron-electron interactions is predicted to lead to charge fractionalization\cite{fractionalcrystal,socfractional1,lubensky,socfractional2,loss1} and, in the presence of superconductors, to parafermions\cite{para1,para2,loss2,loss3}. { In the absence of superconductors, a powerful tool to investigate the emergence of fractionalization phenomena is represented by the study of the density oscillations}: the fractional charge oscillations that emerge can compete with Wigner oscillations, and, eventually, be dominant when Wigner oscillations have a less favorable scaling exponent, or when they are absent\cite{fractionalcrystal,socfractional1}. From a technical point of view, fractional charge oscillations are due to sine-Gordon-type terms in Luttinger liquid Hamiltonians\cite{voit,giamarchi}, and the mathematical treatments which are usually performed rely on strong coupling limits. This issue represents a weakness in comparison to the Luttinger liquid theory of Wigner oscillations. In fact, the onset of Wigner oscillations in one dimensional quantum dots can be very well captured by a Luttinger liquid theory enriched by a first order perturbation theory in umklapp scattering. The results obtained in this context are, in fact, in very good agreement with numerical results\cite{numwig1}.\\
\noindent The aim of this work is twofold: on the one hand we establish the presence of fractional charge oscillations in a quantum spin Hall quantum dot in the presence of two particle backscattering, without relying on strong coupling approximations but by means of a simple perturbative approach. For completeness, we also show that the spin density of the system only acquires small corrections of wavevectors $2k_F$ and $6k_F$, $k_F$ being the Fermi momentum, which are strongly suppressed by their scaling exponents. These corrections do not significantly alter the $2k_F$ spin oscillation characterizing the system in the absence of two-particle backscattering. On the other hand, the system we inspect is characterized by fractional charges induced, via the Jackiw-Rebbi mechanism, by the magnetic barriers defining the quantum dot. It, hence, represents an ideal playground for studying the interplay between the two different kinds of charge fractionalization. In this context we show how the relative magnetization of the barriers affects the charge and spin density profiles.\\
The outline of this work is the following: we start by presenting the model for the quantum spin Hall system, first without any confinement, and then in the quantum dot setup. Particular attention is devoted to the two-particle backscattering term, which is demonstrated to be important even away from half filling. Then, we address the effect of two-particle backscattering on the charge density characterizing the quantum dot setup: we show that fractional oscillations emerge, and that their wavelength depends on the fractional charge trapped in the dot. Finally, we address the spin density oscillations, which also depend, in an interesting way, on the Jackiw-Rebbi charge.
\section{Model and two-particle backscattering}
\label{sec:model}
The main ingredient of our model is a one-dimensional edge of a two-dimensional topological insulator. We fix our reference frame so that spin up/down electrons move right/left. We adopt, for now, periodic boundary conditions on a length $\mathcal{L}$. The Hamiltonian $H_0$ reads\cite{helical1,helical2}
\begin{equation}
H_0=\int_0^{\mathcal{L}}dx\Psi^\dagger(x)(-iv_F\sigma_3\partial_x)\Psi(x),
\end{equation}
where $\Psi(x)=(\psi_+,\psi_-)^T$ is the Fermi spinor, with $\pm$ the spin projection, $v_F$ the Fermi velocity and $\sigma_3$ is the third Pauli matrix in the usual representation. When contact density-density interactions are added, the Hamiltonian becomes a helical Luttinger liquid with Hamiltonian
\begin{equation}
H=\frac{1}{2\pi}\int_0^{\mathcal{L}}dx uK(\partial_x\theta)^2+\frac{u}{K}\left(\partial_x\phi\right)^2+\frac{\pi u}{LK}(N_+^2+N_-^2),
\end{equation}
where $u$ is the velocity of the bosonic excitations, $K$ is the Luttinger parameter ($K<1$ for repulsive interactions, and $K=1$ in the absence of interactions), $\theta$ and $\phi$ are the Luttinger bosonic fields and $N_\pm$ are operators counting spin up/down electrons respectively. In terms of the bosonic fields, the Fermi fields read
\begin{equation}
\psi_{\pm}(x)=\frac{U_\pm}{\sqrt{2\pi\alpha}}e^{\pm i\frac{2\pi N_\pm x}{L}}e^{-i(\pm\phi(x)-\theta(x))},
\end{equation}
where $\alpha$ is the Luttinger liquid cutoff and $U_\pm$ are Klein factors.
When axial spin symmetry is broken, a new interaction term, which becomes relevant in the RG sense for $K<1/2$, can emerge\cite{helical1}, namely two-particle backscattering $H_{2p}$. Explicitly, in the fermionic language, one has
\begin{equation}
H_{2p}=g_{2p}\int_0^L dx \psi^\dag_+(\partial_x\psi^\dagger_+)(\partial_x\psi_-)\psi_-+\mathrm{h.c.},
\end{equation}
where $g_{2p}$ is the coupling constant. The process amounts to flipping two spins, and hence to backscatter two electrons. In order to better understand the physical process involved, it is useful to expand the Fermi operators on the eigenstates of $H_0$ with wavevector $k=2n\pi/\mathcal{L}$, $n$ being an integer. Explicitly, we use
\begin{equation}
\psi_{\pm}(x)=\sum_{k}\frac{e^{ikx}}{\sqrt{\mathcal{L}}}c_{k,\pm}
\end{equation}
with $c_{k,s}$ Fermionic annihilation operators. One finds
\begin{equation}
H_{2p}=\frac{g_{2p}}{\mathcal{L}}\sum_{k_1,k_2,q}k_2(k_2+q)c^\dag_{k_1,+}c^\dag_{k_2,+}c_{k_2+q,-}c_{k_1-q,-}+\mathrm{h.c.}.
\end{equation}
It is important to note that $H_{2p}$ commutes with the total momentum, since it respects translational invariance, while it does not commute with the noninteracting Hamiltonian, just as usual Coulomb interactions. There is however an important difference with respect to usual interactions: when density-density interactions are considered, the terms associated to $q=0$ and $q=2k_F$ (with $k_F$ the Fermi momentum) mix noninteracting levels which are very close in energy, independently of the chemical potential. On the other hand, when the chemical potential $\mu$ is not at the Dirac point, all the virtual transitions that two-particle backscattering can induce couple states which, with respect to the noninteracting Hamiltonian, are at least $4\mu$ apart in energy (see Fig.1(b) and (c)). The importance of two particle backscattering is hence expected to be reduced, but not immediately negligible, as the chemical potential is tuned away from the Dirac point. An intuitive way to convince ourselves that $H_{2p}$ should be taken into account even when the system is away from the Dirac point is to consider the effects of a very similar, although much simpler, operator: a uniform ferromagnetic coupling in the $x$ direction. Specifically, consider the contribution to the Hamiltonian
\begin{equation}
H_B=B\int_0^{\mathcal{L}}dx\psi^\dag_+(x)\psi_-(x)+\mathrm{h.c.}=B\sum_{k}c^{\dag}_{k,+}c_{k,-}+\mathrm{h.c.}.
\end{equation}
Exactly as $H_{2p}$, $H_B$ conserves the total momentum and, when the chemical potential is away from the Dirac point, the energy threshold for virtual states is non-zero (it is $2\mu$ instead of $4\mu$ since only a single electron is now involved). Still, it is very simple to diagonalize the Hamiltonian $H_0+H_B$ and to convince ourselves that all eigenstates and all eigenvectors are modified by $H_B$, although, of course, as the chemical potential is tuned away from the Dirac point, the effects of the magnetic coupling decrease. A scheme of the virtual transitions induced by the ferromagnetic coupling are shown in Fig.1(a).\\
\begin{figure}[htbp]
\begin{center}
\includegraphics[width=8.5cm,keepaspectratio]{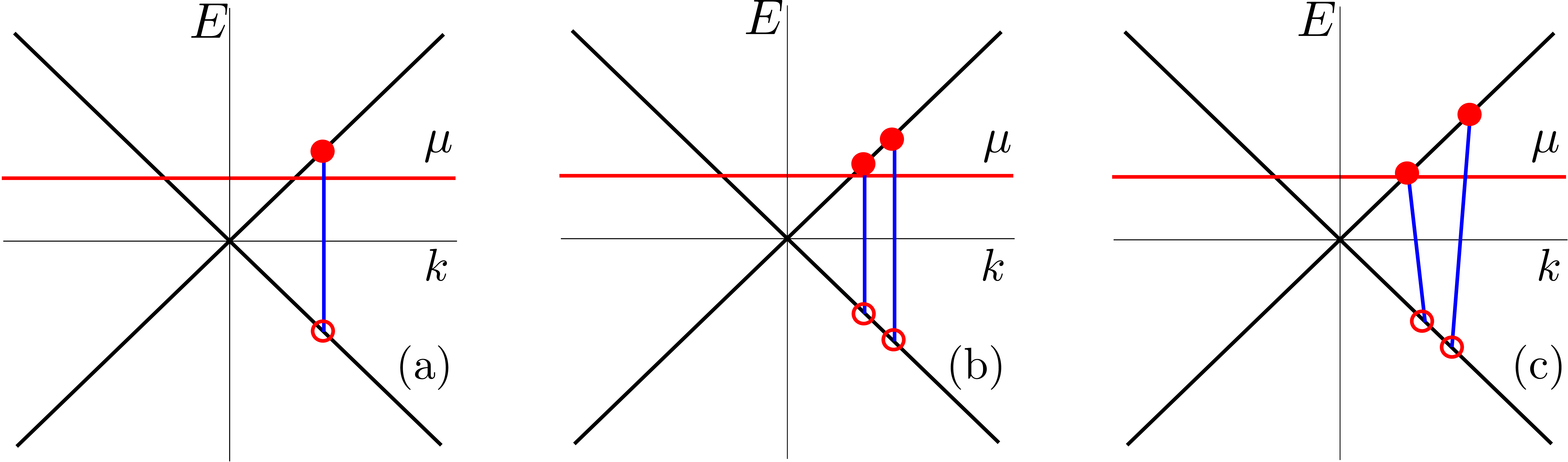}
\caption{(Color online) Schematic of the dispersion relation of a helical one-dimensional system, where the chemical potential $\mu$ is indicated. The virtual transitions associated to (a) magnetic fields, (b) two-particle backscattering at $q=0$, (c) two-particle backscattering for $q\neq 0$ are shown. }
\label{fig:fig1}
\end{center}
\end{figure}
Moreover, it is worth pointing out that two-particle backscattering in helical systems is essentially a spin-spin interaction, since one has
\begin{equation}
H_{2p}\propto g_{2p}\int_0^L dx s_x(x)^2+2\psi^\dagger_+(x)\psi_+(x)\psi^\dagger_-(x)\psi_-(x),
\end{equation}
where $s_x(x)=\psi^\dagger_+(x)\psi_-(x)+\mathrm{h.c.}$, so that, if the system is characterized by a nontrivial spin texture, two-particle backscattering is expected to induce interesting modifications thereof.
\section{The helical quantum dot}
The model we now inspect is an interacting quantum dot built at the edge of a two-dimensional topological insulator. Namely, a small part, of length $L\ll \mathcal{L}$, of the system inspected in the previous section, hosts a quantum dot. The confinement mass is provided by two magnetic barriers, whose magnetization is assumed to lie on the plane of the quantum spin Hall bar. The angular difference between the magnetization of the two barriers is assumed to be $\theta$. For a scheme, see Fig.2.
\begin{figure}[htbp]
\begin{center}
\includegraphics[width=6.5cm,keepaspectratio]{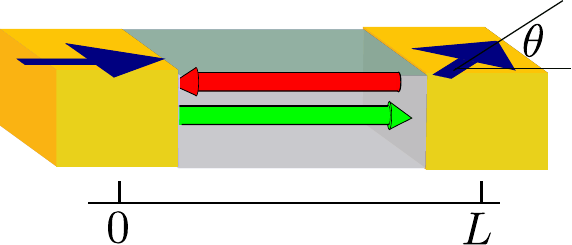}
\caption{(Color online) Schematic of the quantum dot. Two magnetic barriers, whose in plane magnetization differs by an angle $\theta$, confine the helical liquid on the segment $[0,L]$}
\label{fig:fig1}
\end{center}
\end{figure}
As previously discussed in the literature\cite{qi}, a fractional background charge $\theta/2\pi$ is trapped in the dot via the Jackiw-Rebbi mechanism. Moreover, the two components of the spinor, in the dot region, are not independent. They satisfy the boundary conditions $\psi_-(x)=-i\psi_+(-x)$\cite{dolc1} with
\begin{equation}
\psi_+(x):=\frac{U}{\sqrt{2\pi\alpha}}e^{i\frac{\pi x}{L}(N-\frac{1}{2}+\frac{\theta}{2\pi})}e^{i\phi(x)}.
\end{equation}
Note that now a single Klein factor $U$ and a single number operator $N$ are sufficient. The interacting Hamiltonian $H$ of the quantum dot reads
\begin{equation}
H=\frac{\pi v_F}{KL}\sum_{n>0}n a_n^\dagger a_n+\frac{\pi v_F}{2K^2L}\left(N+\frac{\theta}{2\pi}\right)^2+H_{2p},
\end{equation}
with $a_n$ bosonic annihilation operators. Moreover,
\begin{eqnarray}
H_{2p}&=&-\frac{g_{2p}}{2(\pi\alpha)^2}\int_0^L \mathcal{H}_{2p},\\
\mathcal{H}_{2p}&=&\cos\left[\frac{4\pi x}{L}\left(N-\frac{1}{2}+\frac{\theta}{2\pi}\right)-4\varphi(x)-4f(x)\right],\nonumber
\end{eqnarray}
with $\varphi(x)=\left[\phi(-x)-\phi(x)\right]$, $f(x)=\left[\phi(x),\phi(-x)\right]/(4i)$ and
\begin{equation}
\phi(x)\!\!=\!\!\sum_{n>0}\!\!\frac{e^{-\frac{\alpha n \pi}{2L}}}{\sqrt{n}}\!\left[\!\frac{1}{\sqrt{K}}\!\cos\!\left(\frac{n\pi x}{L}\right)\!+\!i\sqrt{K}\!\sin\!\left(\frac{n\pi x}{L}\right)\!\right]\!a_n+\!\mathrm{h.c.}.
\end{equation}
The quantum dot described so far has interesting properties even in the absence of $H_{2p}$: as already mentioned, a fractional charge $\theta/2\pi$ is trapped into it. Moreover, the spin helix characterizing the usual unconfined helical Luttinger liquid describing each edge of a quantum spin Hall system is here pinned by the magnetic impurities. This pinning gives rise to nontrivial spin oscillations, whose typical wave vector $k_s=2\pi x/L(N-1/2+\theta/(2\pi))$ depends on the fractional charge trapped in the system. However, the zero temperature average charge density is flat, due to the absence of Friedel oscillations in the original helical Luttinger liquid. It is worth to mention that, in contrast to spin-orbit coupled one-dimensional quantum dots,\cite{gambetta} there are no states localized at the barriers. In the next sections, we address how two particle backscattering affects the charge and spin densities of the quantum dot.
\section{Particle density}
Due to spin-momentum locking, the density-density correlation functions of the helical Luttinger liquid, in the absence of two-particle backscattering, do not show signatures of Friedel and Wigner oscillations\cite{nof,nof2}. The presence of impurities does not alter this behavior\cite{dolc1}: the density operator $\rho(x)$ reads
\begin{equation}
\rho(x)=\Psi^\dagger(x)\Psi(x)=\frac{N}{L}+\frac{\theta}{2L\pi}-\frac{i\sqrt{K}}{L}\sum_{n>0}\gamma_n (a^\dag_n-a_n)
\end{equation}
with
\begin{equation}
\gamma_n=\sqrt{n}e^{-\frac{n\alpha\pi}{2L}}\cos{\frac{n\pi x}{L}}.
\end{equation}
The expectation value $\bar{\rho}_0(x)$ of $\rho(x)$ on the $N$-electron ground state $|N\rangle$ of $H_0$, is
\begin{equation}
\bar{\rho}_0(x)=\langle N|\rho(x)|N\rangle=\frac{N}{L}+\frac{\theta}{2L\pi}.
\end{equation}
In the absence of two particle backscattering, the ground state average density is hence uniform, although corrected from the simple value $N/L$ by the Jackiw-Rebbi contribution. When two particle backscattering is added, the state $|N\rangle$ is not the ground state of the theory anymore. The first order correction $\delta\bar{\rho}(x)=\bar{\rho}(x)-\bar{\rho}_0(x)$ of the electron density $\bar{\rho}(x)$ averaged over the ground state of $H=H_0+H_{2p}$, reads
\begin{eqnarray}
\delta\bar{\rho}(x)&=&2\mathrm{Re}\sum_{|n\rangle\neq |N\rangle}\frac{\langle N|\rho(x)|n\rangle\langle n| H_{2p}|N\rangle}{E_{|N\rangle}-E_{|n\rangle}}=\nonumber\\
&=&d_0\int_0^L dy \left\{\sin\left[\frac{4\pi y(N+\frac{\theta}{2\pi}-\frac{1}{2})}{L}-4f(y)\right]\right.\nonumber\\
&\cdot&\left.\left(f\left(\frac{x+y}{2}\right)-f\left(\frac{x-y}{2}\right)\right)S(y)^{4K}\right\}\label{eq:den}.
\end{eqnarray}
Here, $d_0={2g_{2p}K^2}/{((\pi\alpha)^2v_F)}$, the states $|n\rangle$ are eigenstates of $H_0$ corresponding to the eigenvalues $E_{|n\rangle}$ and
\begin{equation}
S(x)=\frac{\sinh\left(\frac{\pi\alpha}{2L}\right)}{\sqrt{\sinh^2\left(\frac{\pi\alpha}{2L}\right)+\sin^2\left(\frac{\pi x}{L}\right)}}
\end{equation}
is a damping factor.
\begin{figure}[htbp]
\begin{center}
\includegraphics[width=8.5cm,keepaspectratio]{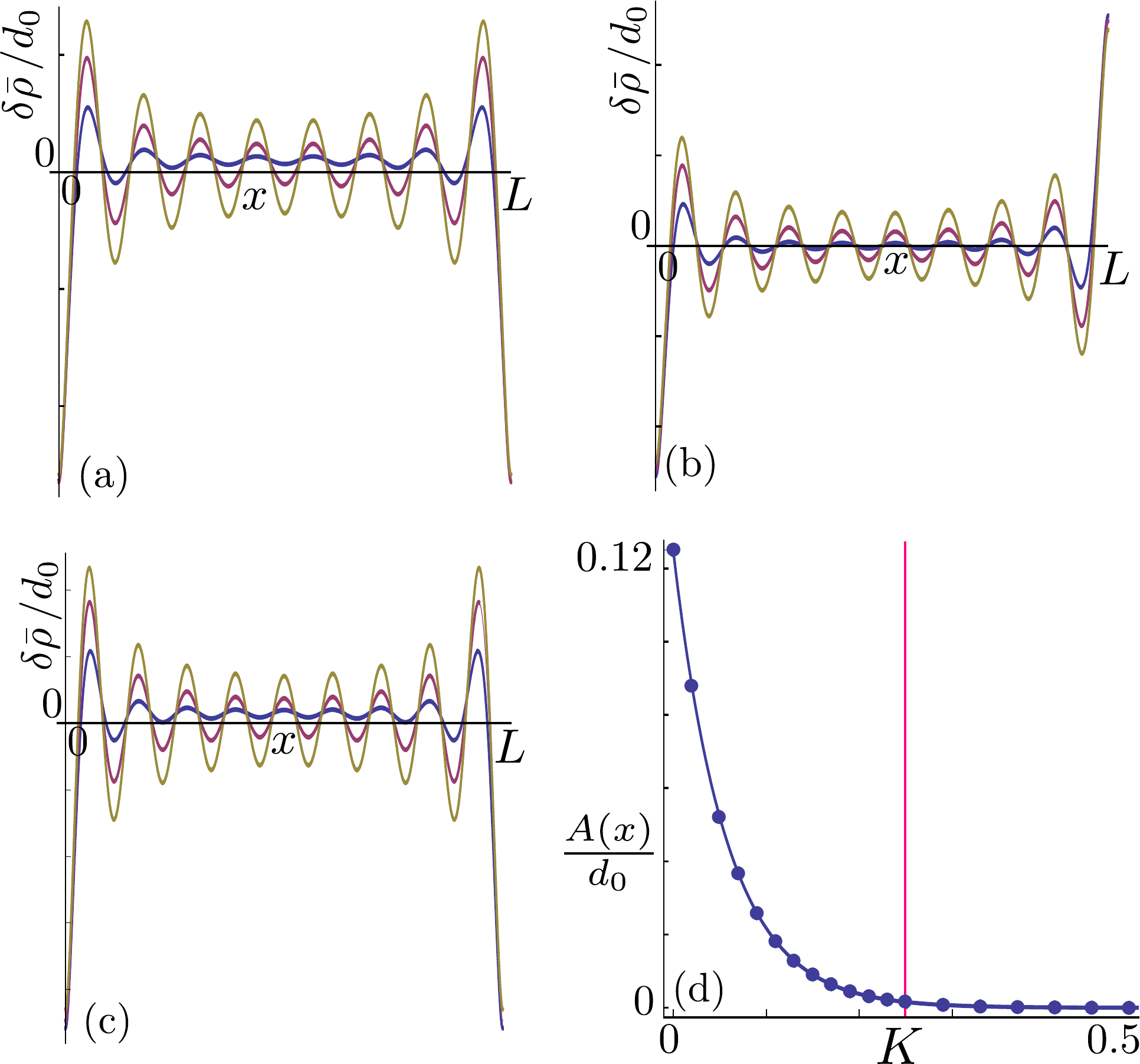}
\caption{(Color online) Two-particle backscattering induced oscillations $\delta\bar{\rho}(x)$ as a function $x$ for $N=4$, $K=0.5$ (brown), $K=0.3$ (violet), $K=0.2$ blue, and (a) $\theta=0$, (b) $\theta=\pi/2$ (c) $\theta=\pi$. (d) The amplitude $A(x)$ of the oscillations as a function of the Luttinger parameter $K$ (dots); the blue line plotted is $A(x)=S(L/2)^{(4K)}$, the pink line is $K=1/4$.}
\label{fig:fig1}
\end{center}
\end{figure}
As interactions are increased ($K$ is reduced), $4k_F$ oscillations emerge, in accordance with the strong coupling limit. Since the system is a one-channel Luttinger liquid, $4k_F$ is not the wave vector of usual Wigner oscillations, which would be $2k_F$\cite{sablikov,gambetta2}. { The presence of dominant $4k_F$ oscillations can therefore be identified as a signature of the emergence of fractional charges in the dot}. Half a period (one maximum of the density) is gained when the magnetic barrier is rotated. The result for different interaction strengths and angles are shown in Fig.3. Although it is beyond the scope of this work, one can speculate that, in the strong coupling regime, when the barrier is is rotated by $\pm\pi$, a sharp wavepacket with fractional charge $e/2$ is introduced/expelled from the quantum dot. This behavior is due to the interplay of the chiral anomaly, responsible for the $\theta$ dependence of the particle number in the dot, and strong interactions. Surprisingly, even in the finite size setup under investigation, the Luther-Emery point\cite{luther} $K=1/4$ plays a crucial role: Intuitively, this point is special because it marks the beginning of the repulsive regime for the effective charge 1/2 fermions\cite{foster}. We have numerically obtained the difference between the relative maximum and the relative minimum of the electron density which are closer to $x=L/2$. This difference, normalized to $d_0$, is referred to as $A(x)$. The numerical points are indicated by the dots in Fig.3(d). The line is proportional to $S(L/2)^{4K}$.  Since in the usual spinless Luttinger liquid the same quantity scales as $S(L/2)^{K}$,\cite{open} the fractional oscillations of the density can be interpreted as Friedel oscillations of the new fermions, which are noninteracting at $K=1/4$. A very drastic simplification of the formula in Eq.(\ref{eq:den}) can be obtained in the limit $\alpha\rightarrow 0$ and $K\rightarrow 0$. Although this limit is outside of the validity of perturbation theory, it allows us to clearly identify the $4k_F$ nature of the oscillation: while the damping factor tends to one, due to the scaling exponent, the term $f((x+y)/2)-f((x-y)/2)$ becomes piecewise linear, and the integral can be easily performed analytically. The correction to the electron density $\delta\bar{\rho}_\infty$, in this limit reads
\begin{equation}
\delta\bar{\rho}_\infty=-d_0\left(\frac{\cos\left[\frac{4\pi x}{L}(N+\frac{\theta}{2\pi})\right]}{\frac{8}{L}\left(N+\frac{\theta}{2\pi}\right)}-\frac{\pi \sin(2 \theta)}{8(2 N \pi + \theta)^2}\right).\label{eq:inf}
\end{equation}
This formula is plotted for clarity in Fig.4 as a function of $x$ and $\theta$. The influence of the chiral anomaly is clear: as $\theta$ is increased, the number of particles and the number of peaks in the density are increased as well. Moreover, the result in Eq.(\ref{eq:inf}) is in qualitative accordance with both the strong coupling regime and physical intuition, since the amplitude of the correction is reduced as the chemical potential is tuned away from the Dirac point.
\begin{figure}
\begin{center}
\includegraphics[width=4.5cm,keepaspectratio]{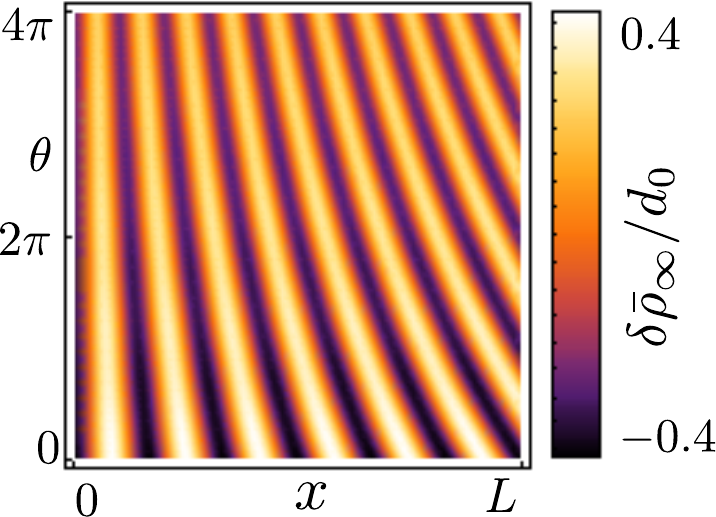}
\caption{(Color online)Density plot of $\delta\bar{\rho}_\infty$, for $N=3$, in units $d_0$ as a function of $x$ and of the angle $\theta$.}
\label{fig:fig1}
\end{center}
\end{figure}
\section{Spin density}
\begin{figure}[htbp]
\begin{center}
\includegraphics[width=8.5cm,keepaspectratio]{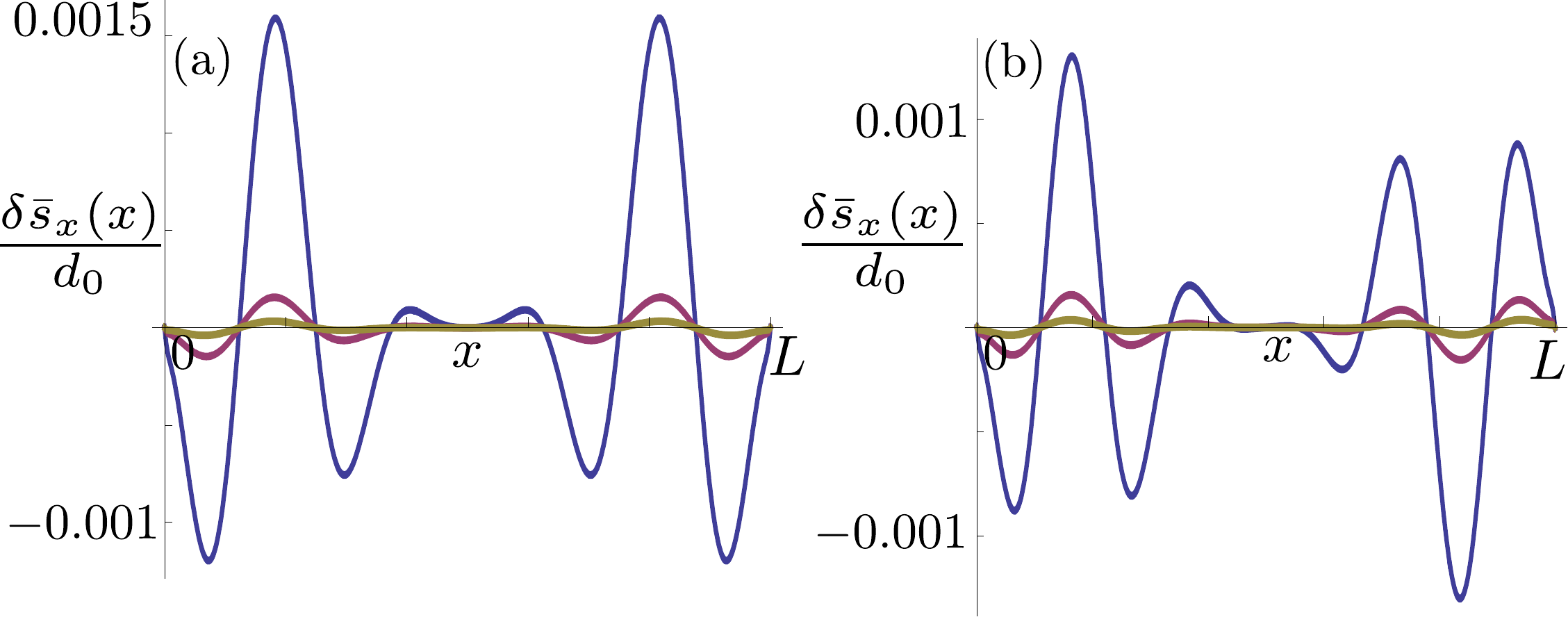}
\caption{(Color online) Two-particle backscattering induced oscillations $\delta\bar{s}_x(x)$ as a function $x$ for $N=4$, $K=0.5$ (brown), $K=0.3$ (violet), $K=0.2$ blue, and (a) $\theta=0$, (b) $\theta=\pi/2$.}
\label{fig:spin}
\end{center}
\end{figure}

In this section, we examine the effect of two particle backscattering on $s_x(x)$ in the quantum dot. The effects on $s_y(x)=\Psi^\dagger(x)\sigma_y\Psi(x)$ are similar, but shifted by half an oscillation, so that the rotating spin pattern discussed in Ref.\cite{dolc2} also characterizes the two-particle backscattering induced corrections. The third component $s_z(x)=\Psi^\dagger(x)\sigma_z\Psi(x)$ is not affected by two-particle backscattering. The correction $\delta \bar{s}_x(x)$ to the average of $s_x(x)$ on the ground state of $H_0$, is given, to first order in $H_{2p}$, by
\begin{equation}
\delta\bar{s}_x(x)=2\mathrm{Re}\sum_{|n\rangle\neq |N\rangle}\frac{\langle N|s_x(x)|n\rangle\langle n| H_{2p}|N\rangle}{E_{|N\rangle}-E_{|n\rangle}}.
\end{equation}
Unfortunately, in the present case we were not able to find an explicit form for the corrections, since the result of the calculation contains entangled series.
Using the well known relations between the matrix elements of the Fermi operator on the Luttinger liquid eigenstates and the Laguerre polynomials\cite{laguerre,grifoni} we could, however, obtain
\begin{equation}
\delta\bar{s}_x(x)=\frac{2g_{2p}}{(2\pi\alpha)^2}\textrm{Re}\int_0^L dy \sum_{\{\mathbf{n}\}\neq\{\mathbf{0}\}}\frac{T(x,y,N,\theta,\{\mathbf{n}\})}{E_{\{\mathbf{n}\}}}.
\end{equation}
Several quantities are here defined: ${\{\mathbf{n}\}\neq\{\mathbf{0}\}}$ is any succession of nonnegative integers. From the physical point of view, the sum emerges from the necessity to consider every possible configuration of the bosonic field, that is, every possible occupation number $n_p$ of the $p$-th bosonic mode. The energy factor is given by $E_{\{\mathbf{n}\}}=\sum_{p=1}^\infty \frac{n_p p v_F}{KL}$. Furthermore, we have that
\begin{eqnarray}
T(x,y,N,\theta,\{\mathbf{n}\})=\sum_{\xi_1=\pm,\xi_2=\pm}\xi_1e^{\frac{2\xi_1 i\pi x(N-1/2+\theta/(2\pi))}{L}-2f(2x)}\nonumber\\
A(x,\xi_1,\{\mathbf{n}\})e^{\frac{4\xi_2 i\pi y(N-1/2+\theta/(2\pi))}{L}-4f(2y)}A(y,2\xi,\{\mathbf{n}\})/(4i),\nonumber
\end{eqnarray}
where
\begin{equation}
A(x,\chi,\{\mathbf{n}\})\!=\!\prod_{p}\frac{\left[-2\chi e^{-\frac{-\alpha p \pi}{2L}}\sqrt{\frac{K}{p}}\sin\left(\frac{p\pi x}{L}\right)\right]^{n_p}}{\sqrt{n_p!}}S(x)^{\chi^2K}.
\end{equation}
The sums and the integrals can be performed numerically and the results are presented in Fig.\ref{fig:spin}.
The dominant oscillation is of $2k_F$ type, although a small $6k_F$ component is present. Moreover, the unfavorable scaling factors of the damping factor $S(x)$ flattens the oscillations near the center of the dot. Additionally, the correction to the spin oscillations is strongly influenced by the Jackiw-Rebbi charge, in accordance with the behavior of the average spin density. As the background fractional charge is added to the dot, the number of oscillations increases. When a full rotation of the magnetization of the barrier is performed, an additional peak is emerging in the spin density profile. As expected, by increasing interactions, the peak-to-valley ratio of spin oscillations increases. However, while the charge oscillations induced by two-particle backscattering add up to a flat original density profile, the spin oscillations are superimposed to a commensurate oscillation pattern.
\section{Conclusions}
\label{sec:conclusions}
In this work, we have inspected the effects of two-particle backscattering on the charge and spin densities of a quantum spin Hall quantum dot. First, we have characterized the different interaction terms. We have shown that, for sufficiently strong interactions, two-particle backscattering must be taken into account even when the system is tuned away from half filling. Then, we have demonstrated, by means of a simple perturbation theory, that the charge density is strongly influenced by two-particle backscattering, since it induces oscillations, with a wavevector that depends on the Jackiw-Rebbi fractional charge trapped in the dot. The peak-to-valley ratio of the oscillations increases as the forward density-density interaction is increased, that is when the Luttinger parameter decreases. An analogous result also holds for the in-plane spin density. More generally, we have shown that the fractional oscillations recently obtained by means of perturbation schemes in the regime of strong two-particle backscattering are a robust effect, which can also be obtained within usual perturbative approaches. {Our work implies that a quantum spin Hall quantum dot displays very rich physics:} interaction-induced fractionalization and Jackiw-Rebbi fractional charges coexist and have a nontrivial interplay.
\begin{acknowledgments}
Financial support by the DFG (SPP1666 and SFB1170 "ToCoTronics"), the Helmholtz Foundation (VITI), and the ENB Graduate school on "Topological Insulators" is gratefully acknowledged.
\end{acknowledgments}

\end{document}